# Localization of electric field distribution in graded core-shell metamaterials

En-Bo Wei[1,2], K. W. Yu[3,4]

[1] Institute of Oceanology, Chinese Academy of Sciences, Qingdao 266071, People's Republic of China. Email: ebwei@ms.qdio.ac.cn

[2] Key Laboratory of Ocean Circulation and Waves, Chinese Academy of Sciences, Qingdao 266071, People's Republic of China.

[3] Department of Physics, The Chinese University of Hong Kong, Shatin, New Territories, Hong Kong, People's Republic of China. Email: kwyu@phys.cuhk.edu.hk

[4] Institute of Theoretical Physics, The Chinese University of Hong Kong, Shatin, New Territories, Hong Kong, People's Republic of China.

The local electric field distribution has been investigated in a core-shell cylindrical metamaterial structure under the illumination of a uniform incident optical field. The structure consists of a homogeneous dielectric core, a shell of graded metal-dielectric metamaterial, embedded in a uniform matrix. In the quasi-static limit, the permittivity of the metamaterial is given by the graded Drude model. The local electric potentials and hence the electric fields have been derived exactly and analytically in terms of hyper-geometric functions. Our results showed that the peak of the electric field inside the cylindrical shell can be confined in a desired position by varying the frequency of the optical field and the parameters of the graded profiles. Thus, by fabricating graded metamaterials, it is possible to control electric field distribution spatially. We offer an intuitive explanation for the gradation-controlled electric field distribution.

## I. Introduction

The topic of electromagnetic cloaking at optical and microwave frequencies has received much attention since the proposed concept of transformation optics by Pendry et al [1] and Leonhardt [2]. It was suggested that invisibility cloaking can in principle be achieved with metamaterials. Schurig et al [3] realized an invisibility cloak experimentally by a core-shell structure and demonstrated the cloaking mechanism in the range of microwave frequencies. Along the similar lines, the spatial coordinate transformations of Maxwell's equation can facilitate the design of a variety of structures with unique electromagnetic or optical functions. Rahm et al [4] designed a square electromagnetic cloak and an omni-directional electromagnetic field concentrator. In fact, a partially resonant mechanism between the induced dipoles and quadrupoles of particles for quasi-static cloaking has been achieved in a region external to the cloaking body by Nicorovici et al [5, 6] and Milton et al [7]. Subsequently, the method was extended by Cui et al [8], Boardman et al [9] and Alu et al [10] to various different cases. Although the experimental realization of cloaking has been imperfect up to now, the concept of a form-invariant coordinate transformation has been applied to variety of electromagnetic, acoustic and elasto-mechanical structures [11-17]. For instance, cloaking in acoustic wave has been observed through the form-invariant acoustic equation [17]. From the above work, it is worthwhile constructing a material whose materials properties can vary gradually in space. We shall call this material the graded metamaterial. Based on graded core-shell metamaterials, it is possible to hide an object from the outside world through altering the flow of light in a graded metamaterial shell. Recently, graded metamaterials for cloaking purpose were designed in terms of spectral representation [18] together with controlling the electric field in cylindrical geometry [19].

Motivated by the above work, we consider the problem of controlling electric field distribution in the case of graded cylindrical core-shell metamaterials. The extension of different graded metamaterial models is related to a few theoretical methods for discussing the effective response and analytical solutions of general graded materials, such as using the special function [20-24], the transformation field method [25] and the differential effective dipole approximation [26-28]. In this paper, with an applied optical field, we consider a coated cylindrical structure, which consists of a homogeneous dielectric core, a shell of graded metamaterial having a complex permittivity given by the graded Drude model [18,29] $\varepsilon_s(r,\omega) = 1 - \omega_p^2(r)/\omega(\omega + i\Gamma)$, where $\omega$ is the angular frequency of the external optical electric field, $\omega_p(r)$ ($\Gamma$) is the spatially varying plasmon frequency (relaxation rate), and $r$ is the radial variable in cylindrical coordinates. For a power-law profile of plasmon frequency $\omega_p(r)$, the potential and electric field distributions can be solved analytically in terms of the hyper-geometric functions. Our results show that the electric field can be controlled by varying the model parameters. Such a control can be achieved by the graded metamaterials at optical or microwave frequencies.

In Sec II, we will establish the general formalism of a graded cylindrical core-shell structure in the quasi-static limit when the optical electric field is applied along the $x$-direction, and obtain the equivalent permittivity of such structure. Furthermore, the analytical potentials are obtained for a power-law profile in a graded Drude model. In Sec.III, based on the analytical solutions, we have numerically investigated how the graded parameters can spatially control the electric field distribution and determine the position of the peak of the field distribution within the two-dimensional core-shell structure. A conclusion will be given in Sec. IV.

## II. Solutions of a graded cylindrical core-shell structure

We suppose the complex permittivity of an isotropic cylindrical medium is a function of its spatial variables, i.e. $\varepsilon_\alpha(r)$ depends on the radial variable $r$ in cylindrical coordinates $(r,\varphi,z)$, and the subscript $\alpha = c, s, h$ notes the core, shell and isotropic host regions, respectively. Note that the coated cylinder has a core of inner radius $a$ and a shell of outer radius $b$. The Maxwell's equations $\nabla \cdot D = 0$ and $\nabla \times E = 0$ give the following field equation in the quasi-static limit,

$$\nabla \cdot [\varepsilon_\alpha(r)\nabla \Phi_\alpha] = 0, \qquad (1)$$

where $\Phi_\alpha$ notes the potential of $\alpha$-type medium and the local electric field $E_\alpha = -\nabla \Phi_\alpha$. For an infinitely long cylinder structure, Eq.(1) can be reduced to the following equation when an external uniform electric field $E_0$ is applied along the $x$-direction,

$$\frac{1}{r}\frac{\partial}{\partial r}[\varepsilon_\alpha(r) r \frac{\partial \Phi_\alpha}{\partial r}] + \frac{1}{r}\frac{\partial}{\partial \varphi}[\varepsilon_\alpha(r)\frac{\partial \Phi_\alpha}{r\partial \varphi}] = 0. \qquad (2)$$

With Eq.(2), the potentials of the homogeneous core and host regions, i.e. constants

$\varepsilon_c$ and $\varepsilon_h$, take the following forms, respectively,

$$\Phi_c(r,\varphi) = ArE_0 \cos(\varphi), \qquad r \leq a \qquad (3)$$

$$\Phi_h(r,\varphi) = (-r + b^2 Br^{-1})E_0 \cos(\varphi). \qquad r > b \qquad (4)$$

For the graded cylindrical shell, the potential $\Phi_s$ can be expressed as $\Phi_s(r,\varphi) = \sum_{m=0}^{\infty} R_m(r)\cos(m\varphi)$, where $R_m(r)$ is the radial function. Substituting the shell potential into Eq.(2), we obtain the equation for the radial function $R_m(r)$,

$$\frac{1}{r}\frac{\partial}{\partial r}[\varepsilon_s(r)r\frac{\partial R_m(r)}{\partial r}] - \frac{m^2}{r^2}[\varepsilon_s(r)R_m(r)] = 0, \quad a < r \leq b. \qquad (5)$$

For Eq.(5), there exist two solutions $R_m^+(r)$ and $R_m^-(r)$ that are regular at origin $r = 0$ and infinity $r \to \infty$, respectively. Then the general solution of the radial function is given by $R_m(r) = A_m R_m^+(r) + B_m R_m^-(r)$. To satisfy the boundary conditions of potentials in Eqs.(3) and (4), the solution in graded cylindrical shell region can be derived

$$\Phi_s(r,\varphi) = [A_1 R_1^+(r) + B_1 R_1^-(r)]E_0 \cos(\varphi), \qquad a < r \leq b. \qquad (6)$$

Considering the continuous boundary conditions of the potential and normal electric displacement at the interfaces at $r = a$ and $r = b$, the non-zero coefficients $A$, $B$, $A_1$ and $B_1$ of the Eq.(3), Eq.(4) and Eq.(6) are determined

$$A = [A_1 R_1^+(a) + B_1 R_1^-(a)]/a = R_1(a)/a, \qquad (7)$$

$$B = \frac{F(b)-1}{F(b)+1}, \qquad (8)$$

$$A_1 = 2bT_2^-(a)/[T_2^+(a)T_1^-(b) - T_1^+(b)T_2^-(a)], \qquad (9)$$

$$B_1 = -2bT_2^+(a)/[T_2^+(a)T_1^-(b) - T_1^+(b)T_2^-(a)], \qquad (10)$$

$$T_1^+(b) = R_1^+(b) + b\frac{\varepsilon_s(b)}{\varepsilon_h}\frac{\partial}{\partial r}R_1^+(b),$$

$$T_1^-(b) = R_1^-(b) + b\frac{\varepsilon_s(b)}{\varepsilon_h}\frac{\partial}{\partial r}R_1^-(b),$$

$$T_2^+(a) = R_1^+(a) - a\frac{\varepsilon_s(a)}{\varepsilon_c}\frac{\partial}{\partial r}R_1^+(a),$$

$$T_2^-(a) = R_1^-(a) - a\frac{\varepsilon_s(a)}{\varepsilon_c}\frac{\partial R_1^-(a)}{\partial r},$$

$$F(b) = \frac{b\varepsilon_s(b)}{\varepsilon_h R_1(b)}\frac{\partial R_1(b)}{\partial r}, \tag{11}$$

where $R_1(a) = A_1 R_1^+(a) + B_1 R_1^-(a)$ and $R_1(b) = A_1 R_1^+(b) + B_1 R_1^-(b)$. Note that $F(b)$ is the equivalent permittivity of the graded cylindrical shell structure and formally dependent on the detailed graded profiles, and $B$ the equivalent dipole factor of the system.

Next, we will consider a radially inhomogeneous permittivity profile given by graded Drude model $\varepsilon_s(r,\omega) = 1 - \omega_p^2(r)/\omega(\omega+i\Gamma)$, where $\omega_p^2(r) = \omega_p^2(0)(1-hr^k)$, and derive the analytic solutions of the graded cylindrical shell region so that a better control of local electric field is discussed. To simply the calculation and without loss of generality, we further normalize the field frequencies $\omega$ and the relaxation rate $\Gamma$ with $\omega_p(0)$ (or let $\omega_p(0)=1$). The normalized Drude model is thus rewritten as

$$\varepsilon_s(r) = 1 - (1-hr^k)/\omega(\omega+i\Gamma). \tag{12}$$

To obtain the analytical solution of the potential of the graded cylindrical shell with gradient profile Eq.(12), we introduce a transformation, namely, $R_1(r) = r^s g(z)$, where $z = hr^k/(\omega^2 + i\omega\Gamma - 1)$, into Eq.(5). For $m=1$, we have

$$k^2 z^2 \frac{\partial^2 g(z)}{\partial z^2} + kz(2s + k - \frac{kz}{1-z})\frac{\partial g(z)}{\partial z} + (s^2 - 1 - \frac{skz}{1-z})g(z) = 0. \tag{13}$$

Furthermore, letting $s^2 = 1$ in Eq.(13), we obtain a hyper-geometric differential equation,

$$z(1-z)\frac{\partial^2 g(z)}{\partial z^2} + [(2s/k + 1) - z(2s/k + 2)]\frac{\partial g(z)}{\partial z} - \frac{s}{k} g(z) = 0. \tag{14}$$

The solution of Eq. (14) can be expressed by the hyper-geometric function $F(\alpha, \beta, \gamma, z)$, which is analytic in the whole complex plane except at singular points $z = 0$, 1 and $\infty$ [30], where the parameters $\alpha$, $\beta$ and $\gamma$ are determined by the following equations, $\gamma = 2s/k + 1$, $\alpha + \beta = 2s/k + 1$ and $\alpha\beta = s/k$. Note that we can demonstrate that the hyper-geometric functions $F(\alpha_{+1}, \beta_{+1}, \gamma_{+1}, z)$ for $s = 1$ and $F(\alpha_{-1}, \beta_{-1}, \gamma_{-1}, z)$ for $s = -1$ are linear independent and construct the general solution of Eq.(14), where $\gamma_{\pm 1} = \pm 2/k + 1$, $\beta_{\pm 1} = \frac{1 \pm 2/k + \sqrt{1 + 4/k^2}}{2}$, $\alpha_{\pm 1} = \frac{1 \pm 2/k - \sqrt{1 + 4/k^2}}{2}$. Thus, in graded cylindrical shell region, we obtain two solutions $R_1^+(r)$ and $R_1^-(r)$, which are again regular at origin point $r = 0$ and infinity $r \to \infty$, respectively,

$$R_1^+(r) = rF(\alpha_{+1}, \beta_{+1}, \gamma_{+1}, hr^k/d), \tag{15}$$

$$R_1^-(r) = r^{-1}F(\alpha_{-1}, \beta_{-1}, \gamma_{-1}, hr^k/d), \tag{16}$$

where $d = \omega^2 + i\omega\Gamma - 1$. Therefore, considering the Eq.(15) and (16) and the coefficients of Eqs.(3), (4) and (6), we have analytically derived the potential solutions of graded cylindrical shell system with a graded Drude model. These solutions can be used to analyze the local electric field distribution.

## III. Numerical results

In order to investigate the local electric field controlled by the model parameters, we first obtain the electric field formulas in the whole system from above analytical potentials. With electric field formula $E_\alpha = -\nabla \Phi_\alpha$, the moduli $|E_h|$ and $|E_s|$ of the local electric fields in the host and shell regions are derived, respectively,

$$|E_h| = E_0[1+(b/r)^4 B^2 + 2B(b/r)^2 \cos(2\varphi)]^{1/2}, \qquad (17)$$

$$|E_s| = E_0[(\partial R_1(r)/\partial r)^2 \cos^2(\varphi) + R_1^2(r)r^{-2}\sin^2(\varphi)]^{1/2}, \qquad (18)$$

where $R_1(r) = A_1 R_1^+(r) + B_1 R_1^-(r)$. Note that the electric field modulus in an inclusion core region is a constant $|E_c| = AE_0$.

With above formulas, the modulus $|E|$ of the corresponding local electric field is calculated and shown in Fig.1, where the spatial distributions of electric field in the coated cylindrical system ($0 \leq r \leq 1.2$) are dependent on the reduced frequency $\omega/\omega_p(0)$ (external electric field $E_0$ is along $x$-direction and let $E_0 = 1$). It is clearly seen that the electric field can be concentrated at a specific position in the shell regions by varying the reduced frequency $\omega/\omega_p(0)$. The annular regions of large intensity (brighter regions) surround the core and are symmetrical about the $y$-axis. Meanwhile, the brighter region shrinks towards the core as the frequency parameter $\omega/\omega_p(0)$ increases. This indicates that controlling and concentrating the electric field is possible by using the core-shell graded metamaterials, where the similar result was obtained for a solid cylindrical inclusion in Ref. [19]. To further examine the electric field distribution in Fig.1, we calculate the electric field along the $x$-axis (i.e. $\varphi = 0$) for different reduced frequency parameters $\omega/\omega_p(0)$ in Fig.2. It can be easily found that, when the frequency parameter increases, the peak position of the electric field modulus shifts toward the core region. Similar to the case of the reduced frequency parameter $\omega/\omega_p(0)$, the effects of the graded parameter $k$ on the electric field are also shown in Fig.3. This can be understood from the fact that the graded Drude model given by Eq.(12) has a decreasing plasmon resonant frequency from $r=a$ to $r=b$ ($0 \leq a < b \leq 1$) with parameters $\omega/\omega_p(0)$ and $k$ increase. In addition, for a given threshold value of electric field modulus, the radial region, whose corresponding electric field value is larger than the threshold value, shrinks when the reduced frequency parameter increases. This intuitively explains why the brightness and its corresponding region decrease with increasing the reduced frequency $\omega/\omega_p(0)$ in Fig.1.

To discuss the influence of the plasmon frequency parameter $h$ on the local electric field distribution, the numerical calculation is taken for various parameter $h$ ranging from 0.1 to 0.9 in Fig.4. Our results show that there is a relatively broad field distribution over the graded metal-shell space for frequency parameter $h$ ranging from 0.34 to 0.4 (similar to the curve $h=0.4$ in Fig.4) and a broad but weaken field distribution for parameter $h$ ranging from 0.1 to 0.3, like the curve $h=0.5$ in Fig.4. This is possibly caused by the surface plasmon resonance band of graded metamaterials leading a broad field distribution [31]. For our example, the frequency parameter $h$ (ranging from 0.34 to 0.4) results in a stronger field distribution over the whole graded shell region comparing with other values of parameter $h$. For the frequency parameter $h=0.7$, 0.8, 0.9, there clearly exists a peak of local electric field as shown in Fig.4. It implies that the electric field can also be controlled by the parameter $h$ and its peak-value of the electric field distribution shifts toward the core due to the Drude model with increasing parameter $h$.

In order to determine the peak's spatial position of electric field distribution in cylindrical shell, we have investigated the frequency resonant condition $\omega/\omega_p(r)=1$ of the external optical field frequency $\omega$ and graded plasmon frequency $\omega_p(r)$ of metamaterial shell. Because the plasmon frequency $\omega_p(r)$ varies spatially, there exists a spatial region of continuous plasmon resonance in the metamaterial shell. If the external field frequency equals a plasmon frequency of metamaterial shell in a spatial point $r$, resonance occurs and induces a strong field in this spatial position. Thus, the local electric field peak's position can be exactly predicted by the frequency resonant condition $\omega/\omega_p(r)=1$. In Fig.5, we have given the exact peak's position predicted by the formula $\omega/\omega_p(r)=1$ and compared with the numerical results obtained from Figs. 2, 3 and 4. Excellent agreement is obtained and the peak's position can indeed be determined exactly. In addition, if $\omega/\omega_p(r)<1$ (or $\omega/\omega_p(r)>1$) in the whole metamaterial shell region, it implies that the shell is the metallic-like material (or dielectric-like material ), and will induce a broad distribution of local electric field (for the case of metallic-like material, see Fig.4 for $h=0.4$, 0.5) due to the salient properties of graded metamaterials [29].

## IV. Conclusions

In summary, the control of the electric field by graded meatmaterials has been investigated. With a graded Drude model, the potentials and electric fields are analytically derived by means of hyper-geometric function for a graded core-shell cylindrical structure. Our results show that the electric field distribution can be controlled by tuning the frequency, the plasmon frequency and graded parameters. One can further confine the field peak position in the shell region by using the surface plasmon resonant condition $\omega/\omega_p(r)=1$. The precise and flexible positioning of the enhanced field may have many applications in optical design, for example, in displaying and printing technologies. Furthermore, it is also instructive to extend our method to the three-dimensional spherical core-shell system for various graded profile by using established analytical solutions of graded spherical composites [20-23], the differential effective dipole approximation [26-28] and the transformation field method for complex structures [25].

## Acknowledgments

This work was supported by the NSFC (Grant No. 40876094), National 863 Project of China (Grant No. 2008AA09A403) and the RGC General Research Fund of the Hong Kong SAR Government.

**Figure captions**

**Fig.1.** The spatial distributions of electric field modulus $|E|/E_0$ in the core-shell graded cylindrical system are displayed for different reduced frequencies $\omega/\omega_p(0) = 0.6$ (top panel), $0.7$ (middle panel), $0.8$ (bottom panel), respectively, where the parameters read $k = 0.499$, $\Gamma/\omega_p(0) = 0.02$, $h = 0.7$, $a = 0.4$, $b = 1$, $\varepsilon_c = 20$, $\varepsilon_h = 1$, and the host region is shown from $r = 1$ to $r = 1.2$ only. Note that the brighter region represents a stronger field.

**Fig.2.** The distribution of electric field modulus $|E|/E_0$ plotted against the reduced frequency parameter $w = \omega/\omega_p(0)$ along the $x$-direction (i.e. $\varphi = 0$), where $k = 0.499$, $\Gamma/\omega_p(0) = 0.02$, $h = 0.9$, $a = 0.2$, $b = 1$, $\varepsilon_c = 20$, $\varepsilon_h = 1$ and the host region is shown from $r = 1$ to $r = 1.2$ only.

**Fig.3.** The distribution of electric field modulus $|E|/E_0$ plotted against the parameter $k$ of graded complex permittivity along the $x$-direction (i.e. $\varphi = 0$), where the other parameters are given in the legend of Fig.2 except for $k$, $h = 0.8$ and $\omega/\omega_p(0) = 0.6$.

**Fig.4**. The distribution of electric field modulus $|E|/E_0$ plotted against the parameter $h$ of graded complex permittivity along the $x$-direction (i.e. $\varphi = 0$), where the other parameters are given in the legend of Fig.2 except for $h$ and $\omega/\omega_p(0) = 0.6$.

**Fig.5** The plot shows the peak's position $r$ determined by the resonance condition $\omega/\omega_p(r) = 1$ for different frequency parameters $w = \omega/\omega_p(0)$ and graded parameters, where the graded plasmon frequency is $\omega_p^2(r) = \omega_p^2(0)(1 - hr^k)$. The exact peak's position $r$ is compared favorably with the numerical peak's position obtained in Figs. 2, 3 and 4.

Fig. 1:

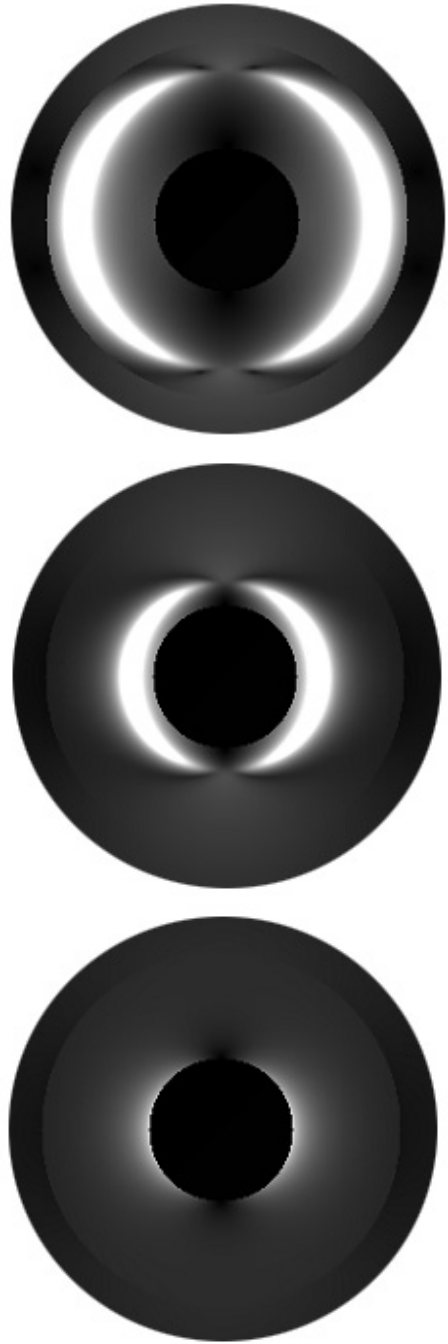

Fig. 2:

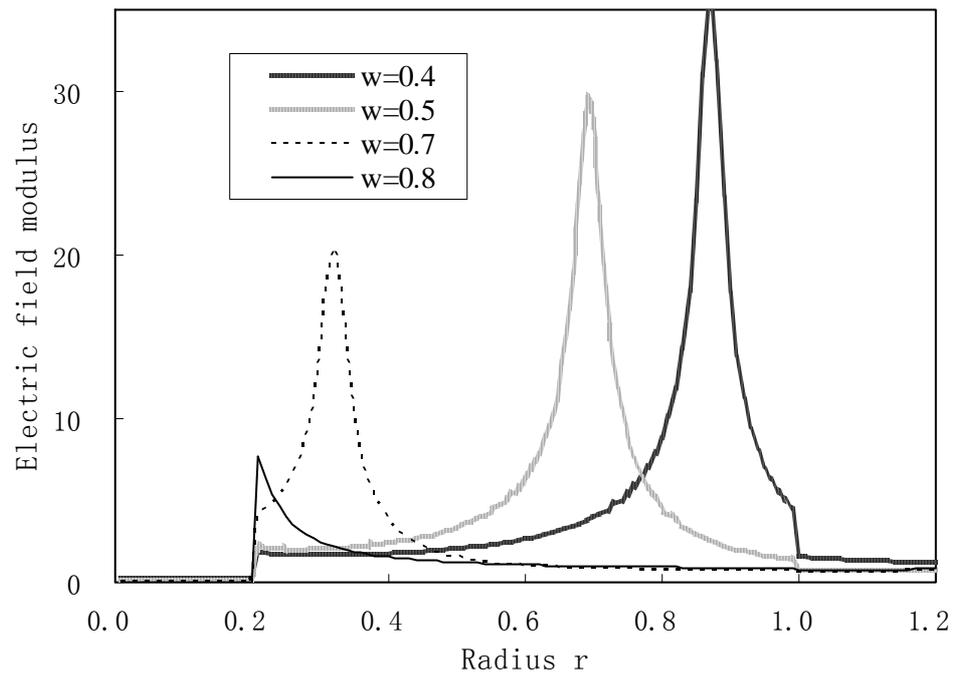

Fig. 3:

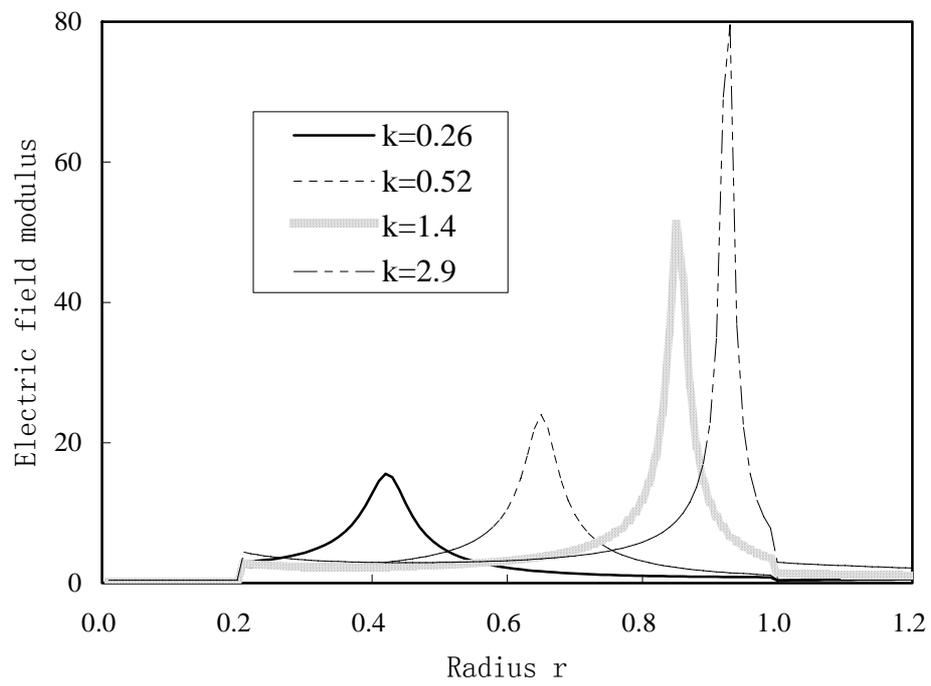

**Fig. 4:**

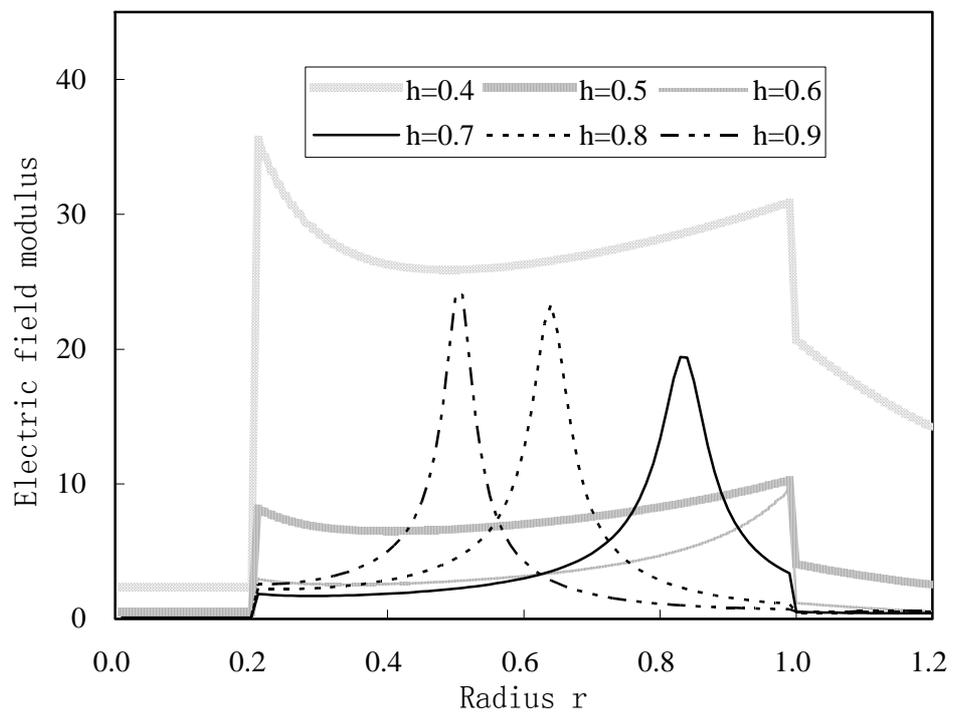

**Fig.5**

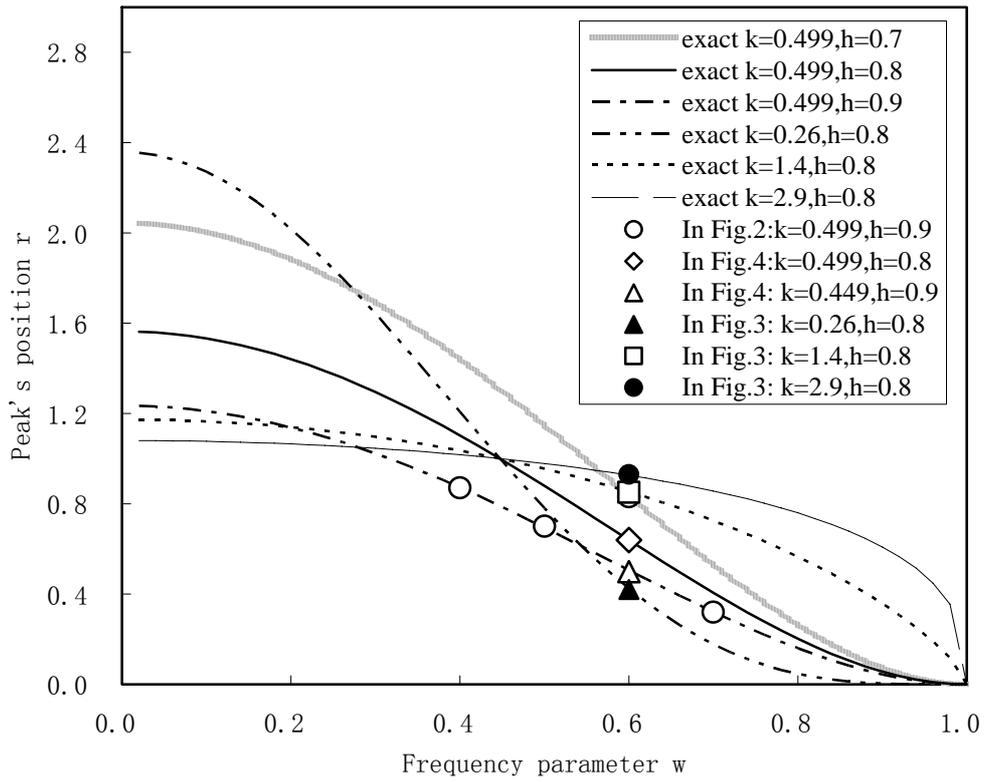